\documentstyle[12pt,world_sci]{article}
\input BoxedEPS.tex         
 \SetRokickiEPSFSpecial      
 \HideDisplacementBoxes      
\newcommand{\be}{\begin{equation}}
\newcommand{\ee}{\end{equation}}
\newcommand{\bdm}{\begin{displaymath}}
\newcommand{\edm}{\end{displaymath}}

\def\fracerr#1{\ifmmode
                    \frac{\delta #1}{#1}
               \else
                    \mbox{${\delta #1}/{#1}$}
               \fi}

\def\D0{D\O }

\def\GeV{\ifmmode
                   {\rm \ GeV}
               \else
                    \mbox{${\rm GeV}$}
               \fi}
\def\TeV{\ifmmode
                   {\rm \ TeV}
               \else
                    \mbox{${\rm TeV}$}
               \fi}
\def\invpb{\ifmmode
                   {{\rm \ pb}^{-1}}
               \else
                    \mbox{${\rm pb}^{-1}$}
               \fi}

\def\Missing#1#2{{\mbox{$#1\kern-0.57em\raise0.19ex\hbox{/}_{#2}$}}}

\def\vMissing#1#2{\ifmmode
            \vec{#1}\kern-0.57em\raise.19ex\hbox{/}_{#2}
         \else
            {{\mbox{$\vec{#1}\kern-0.57em\raise.19ex\hbox{/}_{#2}$}}\ }
         \fi}

\def\St{\ifhmode {$S_T$\ }\else{S_T}\fi}

\def\zino#1{\ifmmode
                {\tilde Z_{#1}}
            \else
                {$\tilde Z_{#1}$}
            \fi}

\def\wino#1{\ifmmode
                {\tilde W_{#1}}
            \else
                {$\tilde W_{#1}$}
            \fi}

\def\squark{\ifmmode
                {\tilde q}
            \else
                {$\tilde q$}
            \fi}

\def\gluino{\ifmmode
                {\tilde g}
            \else
                {$\tilde g$}
            \fi}
\begin{document}
\title{{\bf A STUDY OF EVENTS WITH LARGE TOTAL TRANSVERSE ENERGY PRODUCED
            IN PROTON-ANTIPROTON COLLISIONS AT $\sqrt{s} = 1.8~TeV$ }}

  \author{ HENRYK PIEKARZ\thanks{Representing the D\O\
Collaboration.}\\
{\em  Florida State University, Tallahassee,\\
Florida, 32306, U.S.A.}\\}
\maketitle
\setlength{\baselineskip}{2.6ex}

\begin{center}
\parbox{13.0cm}
{\begin{center} ABSTRACT \end{center}

{\small \hspace*{0.3cm}
Properties of events originating from
proton--antiproton interactions in which the total transverse energy,
$\Sigma\vert E_T \vert$, of the event exceeded
$400$ $GeV$ are presented. These events were produced at the Fermilab
Tevatron Collider operating at a center--of--mass energy of
$1.8~TeV$ and recorded in the D\O\ detector.  We describe our analysis method
which minimizes the effect of multiple interactions in the data sample.
Based on a data sample of $5.45 \pm 0.65~pb^{-1}$, the topology of
these hard scattering events as well as
preliminary results for the cross--section,
$d\sigma/d\Sigma\vert E_T \vert$, are presented and discussed.
}}
\end{center}
\vspace{1.0cm}

The motivation for studying events with large $\Sigma\vert E_T \vert$ is
threefold.  First, the phenomenology of events with the highest
$\Sigma\vert E_T \vert$ is of great interest as such events result from the
hardest scatterings, the most central collisions.  Second, these events
allow us to test our understanding of QCD
where predictions are extrapolated from
lower energy data.  Third, these events provide a window
on physics beyond the Standard Model.  One such scenario is
quark compositeness, characterized by a contact interaction\cite{L1} with
energy scale $\Lambda^{^*}$, which
could be seen as an enhancement
in the cross--section, $d\sigma/d\Sigma\vert E_T \vert$, at large values of
$\Sigma\vert E_T \vert$.  In Fig. 1 we compare
$d\sigma/d\Sigma\vert E_T \vert$ for
QCD (no compositeness) with QCD (including compositeness with
$\Lambda^{^*} = 1.5~TeV$).  These curves
were generated using PYTHIA V5.6 (including the quark
compositeness model of Ref. 1) and
the D\O GEANT detector simulation.
They show the sensitivity at leading order
to the existence of composite quarks at a scale of $\Lambda^{^*} = 1.5~TeV$
for $\Sigma\vert E_T \vert > 500~GeV$.

\begin{figure}[htbp]
\center{ \mbox{{\ForceWidth{7.0 cm} \TrimBottom{6.0 cm} \TrimLeft{2.5 cm}
         \TrimTop{6.0 cm}
         \hSlide{0.6 cm} \vSlide{-0.5 cm} \BoxedEPSF{dpf_talk_fig1a.ps}}}
       }
\caption{~~PYTHIA generated
$d\sigma/d\Sigma\vert E_T \vert$ for
QCD (no compositeness -- dotted line) and QCD
(compositeness at $\Lambda^{^*} = 1.5~TeV$ -- solid line)
indicating the sensitivity at leading order
of $\Sigma\vert E_T \vert$ to quark compositeness.
}
\label{Fig. 1}
\end{figure}

The data presented here were obtained using the D\O\ detector\cite{L3} at
the Fermilab Tevatron Collider at $\sqrt{s} = 1.8~TeV$.  They
were collected in Run 1a, from March -- May, 1993 and correspond
to an integrated luminosity of $5.45 \pm 0.65~pb^{-1}$. The D\O\ detector has a
hermetic, compensating sampling calorimeter with fine longitudinal and
transverse segmentation in azimuth, $\phi$, and pseudorapidity,
$|\eta| \leq 4.2$.
The calorimeter has good energy resolution which can be parametrized as
$\sigma/E = A/(\sqrt{E}) (E~{\rm in}~GeV)$, where
$A=0.15$ for electrons and 0.50 for single hadrons.

The hardware trigger used a scalar sum of $E_T$ in calorimeter cells
in the range, $|\eta|\leq 3.2$ with a threshold on the sum of $225~GeV$.  The
software trigger increased this threshold to $300~GeV$ and imposed an upper
limit on the total energy ($E_{tot}$) for the event of $1.8~TeV$, both
calculated over the range, $|\eta| \leq 4.0$.  By requiring only a minimum
$E_T$
in the calorimeter we are naturally sensitive to multiple
interaction events where a less central collision plus several minimum bias
events will satisfy the trigger conditions.
The cut on $E_{tot}$ was necessary to limit
the rising trigger rate due to multiple interactions as the luminosity
increased.

Following event reconstruction the data sample was further reduced by
increasing
the $\Sigma\vert E_T \vert$ threshold to $370~GeV$ and
applying the same $E_{tot}$ cut of $1.8~TeV$ used in the Level 2 trigger.
These cuts were applied to reconstructed quantities over the range,
$\vert\eta\vert \leq 4.0$.  In addition, we removed events with ``bad'' jets,
which are defined as resulting from noisy calorimeter cells or accelerator
Main Ring activity, and events with isolated high-$E_T$ electron candidates.
Events with multiple interactions occurring in the detector remain in the
sample, and the mean $\Sigma\vert E_T \vert$ rises with the number of
interactions present.  To obtain a variable which does not show a large change
with the addition of extra minimum bias events, we redefine
$\Sigma\vert E_T \vert$ to be a sum over energy clusters only; that is, we sum
over all NN jets found in the event.
Jets were found using fixed cone algorithms
as well as a nearest neighbor (NN) algorithm.  The NN algorithm starts with a
seed tower $E_T$ of $0.5~GeV$ and searches for all towers within a radius,
${\rm R} = \sqrt{(\Delta\eta)^2 + (\Delta\phi)^2} \leq 0.2$, above an
$E_T$ threshold
of $0.5~GeV$.  It continues to sum all such towers that it finds; the
resultant energy cluster is termed a jet if it has $E_T > 8~GeV$.  The NN jet
algorithm, by requiring almost contiguous towers, is
much less
sensitive to both energy fluctuations in the calorimeter, which provide a
seed tower for jet finding,
and to extra minimum bias events.  Using our PYTHIA--generated event sample
we find that the ratio
of $\Sigma\vert E_T \vert (clusters)$ to $\Sigma\vert E_T \vert (cells)$ is
$\sim 0.95$.  That is, $\sim 95 \%$ of the $\Sigma\vert E_T \vert$ of the hard
scattering event is contained in $\Sigma\vert E_T \vert (clusters)$.
In addition, we find that $\Sigma\vert E_T \vert (clusters)$ increases only
minimally as minimum bias events are added to each event.  This very weak
dependence of $\Sigma\vert E_T \vert (clusters)$ on additional minimum bias
events overlaying the hard scattering event means that we can still
extract useful information about the hard scattering event even in the multiple
interaction environment.  To select only the hardest
scattering events we require
$\Sigma\vert E_T \vert (clusters) \geq 400~GeV$.

The efficiency for both the trigger and the offline data
reduction was studied as a function of $\Sigma\vert E_T \vert (clusters)$.  For
$\Sigma\vert E_T \vert (clusters) \geq 400~GeV$ we find this efficiency
to be $> 99\%$.  The
efficiency of the $E_{tot}$ cut is dependent on both the instantaneous
luminosity as well as the $\Sigma\vert E_T \vert (clusters)$.
We express this as a survival fraction which is flat for
$\Sigma\vert E_T \vert (clusters) < \sim 600~GeV$ and then linearly decreasing
above $600~GeV$.  As the instantaneous luminosity increases, the survival
fraction in the flat region decreases, and the linear decrease above that
becomes steeper.  This dependence is exactly what one would hypothesize for
increasing luminosity as the number of multiple interactions also increases.
For NN jets, the removal of ``bad'' jets as defined previously has
an efficiency of $\sim 96\%$, nearly independent of $E_T$,
and is applied on an
event by event basis since it depends on the number of jets.
Other corrections applied to the data concern
the energy determination.  The effect of
multiple interactions on $\Sigma\vert E_T \vert (clusters)$
has been estimated to be $< 2\%$ for clusters formed using NN jets.
Jet energies have been corrected using the D\O\ standard
corrections for cone 0.5 jets.  This correction is applied to the
hadronic energy of the jet and has a magnitude of $\sim 20 \%$; it
underestimates the NN jet corrections by
$\sim 2-4 \%$ and is still under study.  The overall energy scale systematic
uncertainty is estimated to be $\sim 5 \%$.

The data we present are based on an integrated luminosity of
$5.45 \pm 0.65~pb^{-1}$.  We have studied the topology of
these events using the following
distributions: inclusive jet $E_T$, $\Sigma\vert E_T \vert (clusters)$, jet
multiplicity, and inclusive jet $\eta$.
As seen in Fig. 2, these events arise from very hard,
central collisions as evidenced by the inclusive jet $\eta$ distribution which
is very strongly peaked at $\eta = 0$ and has very few entries for
$\vert\eta\vert > 2$.  The inclusive jet $E_T$ distribution shows a peak at
about $200~GeV$.  Combined with the jet multiplicity distribution showing
that most of our events have 2, 3, or 4 jets, this is easily explained as
a result of
our threshold of $400~GeV$ for $\Sigma\vert E_T \vert (clusters)$ where we
have two very stiff jets and one or two small jets accompanying them.
We observe about $14~{\rm events}/pb^{-1}$ for
$\Sigma\vert E_T \vert (clusters) \geq 500~GeV$.
Applying the corrections for inefficiencies and energy scale
described above, we produce a preliminary cross--section,
$d\sigma/d\Sigma\vert E_T \vert$, which is also shown in Fig. 2.


Our highest
$\Sigma\vert E_T \vert (clusters)$
event has $\Sigma\vert E_T \vert (clusters) = 830.6~GeV$ and
a calorimeter
missing $E_T = 10.8~GeV$.  It is a 3 NN jet event with all of the jets at very
central $\eta$.  The event has
$\Sigma\vert E_T \vert (cells) = 920.9~GeV$ and $E_{tot} = 1445.9~GeV$.
Analyzing the data from the Level 0 trigger counters and the number of vertices
found by the central tracker indicates that this is probably a multiple
interaction event.  Based on the very weak dependence of
$\Sigma\vert E_T \vert (clusters)$ on multiple interactions, we expect the
$\Sigma\vert E_T \vert$ of this hard scattering event to be $830.6~GeV$.

We have presented results of an analysis of the hardest scattering events
detected in the D\O\ detector.  We have argued that
$\Sigma\vert E_T \vert (clusters)$ is the correct variable
for selection and study
of these events since it is only minimally affected by multiple interactions.
We have also shown
that $\Sigma\vert E_T \vert (clusters)$, at least in leading order
calculations, provides sensitivity to
new physics, like quark compositeness.  With an order of
magnitude increase in the integrated luminosity, we hope to probe
the scale, $\Lambda^{^*} = 1.5~TeV$, and may even be
able to push it higher.

\begin{figure}[htbp]
\center{ \mbox{{
         \ForceWidth{7.0 cm} \TrimBottom{6.0 cm} \TrimLeft{2.5 cm}
         \TrimTop{6.0 cm}
         \hSlide{0.6 cm} \vSlide{-0.5 cm} \BoxedEPSF{dpf_talk_fig3a.ps}
         \ForceWidth{7.0 cm} \TrimBottom{6.0 cm} \TrimLeft{2.5 cm}
         \TrimTop{6.0 cm}
         \hSlide{0.0 cm} \vSlide{-0.5 cm} \BoxedEPSF{dpf_talk_fig3b.ps}}}
       }
\center{ \mbox{{
         \ForceWidth{7.0 cm} \TrimBottom{6.0 cm} \TrimLeft{2.5 cm}
         \TrimTop{6.0 cm}
         \hSlide{0.6 cm} \vSlide{-0.5 cm} \BoxedEPSF{dpf_talk_fig3c.ps}
         \ForceWidth{7.0 cm} \TrimBottom{6.0 cm} \TrimLeft{2.5 cm}
         \TrimTop{6.0 cm}
         \hSlide{0.0 cm} \vSlide{-0.5 cm} \BoxedEPSF{dpf_talk_fig1b.ps}}}
       }
\caption{
{}~~Kinematic distributions of events with
$\Sigma\vert E_T \vert (clusters) \geq 400~GeV$ and
preliminary cross--section, $d\sigma/d\Sigma\vert E_T \vert$,
based on $5.45 \pm 0.65~pb^{-1}$ of data.}
\label{Fig. 2}
\end{figure}

\vfill\eject
\end{document}